\documentclass{article}
\usepackage{spconf,amsmath,graphicx}
\usepackage{amssymb}
\usepackage{float}
\usepackage{algorithm}
\usepackage{algorithmic}

\def\L{{\cal L}}
\def\P{{\rm P}}
\def\G{{\cal G}}
\def\A{{\cal A}}
\def\R{{\mathbb{R}}}
\newcommand{\nmat}[1]{\boldsymbol{\mathrm{#1}}}
\newcommand{\nvec}[1]{\boldsymbol{#1}}
\newcommand{\assoc}[1]{\mathrm{Assoc}_{#1}}
\newcommand{\cut}[1]{\mathrm{Cut}_{#1}}
\newcommand{\Tr}{\mathrm{Tr}}
\newcommand{\minus}{\scalebox{0.75}[1.0]{$-$}}

\title{Constrained Spectral Clustering for Dynamic Community Detection}
\name{Abdullah Karaaslanli and Selin Aviyente \thanks{This work was in part supported by NSF CCF-1422262.}}
\address{Department of Electrical and Computer Engineering, Michigan State University, East Lansing, MI.}
\begin{document}
\ninept
\maketitle


\begin{abstract}
Networks are useful representations of many systems with interacting entities, such as social, biological and physical systems. Characterizing the meso-scale organization, i.e. the community structure, is an important problem in network science. Community detection aims to partition the network into sets of nodes that are densely connected internally but sparsely connected to other dense sets of nodes. Current work on community detection mostly focuses on static networks. However, many real world networks are dynamic, i.e. their structure and properties change with time, requiring methods for dynamic community detection. In this paper, we propose a new stochastic block model (SBM) for modeling the evolution of community membership. Unlike existing SBMs, the proposed model allows  each community to evolve at a different rate. This new model is used to derive a maximum a posteriori estimator for community detection, which can be written as a constrained spectral clustering problem. In particular, the transition probabilities for each community modify the graph adjacency matrix at each time point. This formulation provides a relationship between statistical network inference and spectral clustering for dynamic networks. The proposed method is evaluated on both simulated and real dynamic networks. 
\end{abstract}


\section{Introduction}
\label{sec:intro}
Community detection (CD) partitions the nodes of a network such that nodes are densely connected  within their respective communities while being sparsely connected across communities \cite{fortunato_community_2010}. CD has important applications in recommendation systems \cite{reddy2002graph}, social networks \cite{moody2003structural} and brain connectomics \cite{sporns_modular_2016}. Recently, CD methods have been developed for networks that change with time, i.e. \textit{dynamic networks} \cite{holme_temporal_2012}. Compared to \textit{static networks}, CD methods for dynamic networks aim to partition nodes at each time as well as to track the changes in the partitions over time \cite{rossetti_community_2018}.

CD in static networks is commonly formulated as the optimization of a quality function. Some of the well-known quality functions are modularity \cite{newman_finding_2004}, normalized and ratio cuts \cite{von_luxburg_tutorial_2007}, InfoMap \cite{rosvall2008maps} and likelihood or posterior distributions defined based on statistical inference \cite{snijders_estimation_1997, karrer_stochastic_2011}. These functions can be divided into two categories \cite{pamfil_relating_2018}. The first category includes functions that are defined heuristically such as modularity or cut based methods. Functions in the second category are based on statistical network models, e.g. \textit{stochastic block model (SBM)} and \textit{degree-corrected SBM (DCSBM)}, and likelihood or posterior distributions are defined as quality functions. 

Dynamic CD methods are mostly based on extensions of aforementioned quality functions from the static to the dynamic case. The early work in dynamic CD named \textit{evolutionary spectral clustering (EvoSC)} \cite{chi_evolutionary_2007}, defines a quality function at each time point as $\alpha CS + \beta CT$ where $CS$ is \textit{snapshot cost}, $CT$ is \textit{temporal cost} and $\alpha,\ \beta$ are parameters that weigh the two terms. This formulation can be thought of as constraining the snapshot cost of a time point with community structure of previous time points. Similarly, modularity optimization \cite{mucha_community_2010, pamfil_relating_2018}, statistical methods \cite{yang_detecting_2011, xu_dynamic_2014, ghasemian_detectability_2016} and InfoMap \cite{peixoto_modelling_2017} have been extended to dynamic networks. 

Recently, there have been attempts to show that heuristic based optimization methods are equivalent to statistical inference under some conditions. Newman et al. \cite{newman_spectral_2013} showed that spectral approximation of modularity, normalized cut and statistical inference are equivalent to each other for a particular choice of parameters. Similarly, equivalence between spectral clustering, modularity maximization and  non-negative matrix factorization is shown in \cite{ma_semi-supervised_2018}. Lastly, in \cite{young_universality_2018} statistical inference is shown to be universal, which means most of the quality functions developed for CD are indeed special cases of statistical inference. This work is extended to dynamic networks in \cite{pamfil_relating_2018}, where dynamic modularity function defined in \cite{mucha_community_2010} is shown to be equivalent to statistical inference methods.

Following this line of work, we propose a method for dynamic CD, referred to as \textit{constrained dynamic spectral clustering (CDSC)}. We start by defining a dynamic DCSBM and the corresponding posterior distribution. We then show that maximizing the posterior distribution can be solved by a dynamic spectral clustering algorithm. The proposed work makes some significant contributions to the literature. First, the proposed dynamic DCSBM allows each community to evolve with a different probability that varies with time. Second, we derive a relationship between statistical inference based on the proposed dynamic DCSBM and spectral clustering, in particular constrained spectral clustering. Finally, we show that the proposed method is a generalization of EvoSC. 

The remainder of the paper is organized as follows. In Section \ref{sec:background}, we give an overview of the notations used in the paper along with a background on spectral clustering and DCSBM. In Section \ref{sec:method}, we introduce our new dynamic DCSBM and the corresponding optimization problem. In Section \ref{sec:results}, the comparison of the proposed method with state-of-the-art dynamic CD methods on both simulated and a real  dynamic network is given. 

\vspace{-1em}
\section{Background}
\label{sec:background}

\vspace{-0.5em}
\subsection{Notation}
A static graph is represented by $G=(V,E)$ where $V$ is the node set with $|V|=n$ and $E\in V\times V$ is the edge set with $|E|=m$. An edge between two nodes $i$ and $j$ is indicated by $e_{ij}$. In this work, the graphs are assumed to be \textit{undirected}, i.e. $e_{ij} = e_{ji}$, and self loops are not allowed, i.e. $e_{ii}\not\in E,\ \forall i\in V$. Each edge $e_{ij}$ is associated with a weight $w_{ij}$. If $w_{ij}\in\{0,1\}$, the graph is said to be \textit{binary}, on the other hand if $w_{}\in\R_{\geq 0}$ then it is a \textit{weighted} graph. \textit{Degree} of a node $i$ is $d_i=\sum_{j}w_{ij}$. A graph is algebraically represented by an $n\times n$ \textit{adjacency matrix} $\nmat{A}$ whose entities are $A_{ij}=w_{ij}$. Lastly, \textit{Laplacian matrix} of a graph is defined as $\nmat{L}=\nmat{D}-\nmat{A}$, where $\nmat{D}$ is a diagonal matrix with entries $D_{ii}=d_i$. 

A dynamic graph is a time sequence of static graphs, i.e. $\mathcal{G}=\{G^1, G^{2}, \dots, G^{T}\}$, where $G^t$s are defined on the same vertex set $V$. The edge sets $E^{t}$s define the set of interactions between nodes at time $t$ \cite{holme_temporal_2012}. Mathematically, we represent $\G$ as a sequence of adjacency matrices $\A = \{\nmat{A}^1, \nmat{A}^2, \dots, \nmat{A}^T\}$.

\vspace{-0.5em}
\subsection{Spectral Clustering}
\label{ssec:spectralclustering}
CD on a graph $G=(V, E)$ is the task of partitioning nodes in $V$ into $K$ non-overlapping communities, i.e. $P = \{C_1, \dots, C_K\}$ where  $C_i\cap C_j = \varnothing\, \forall, i\neq j$ and $\bigcup_{i=1}^K C_i = V$. This task is usually achieved by optimizing a function that quantifies the quality of the communities, $C_i$s. Two widely used quality functions are graph cut and graph association, which are defined as follows. Let  $\nvec{g}$ be  the $n$ dimensional \textit{community assignment vector} whose entries $g_i=k$ if node $i$ is in community $k$ and $\nvec{Z} \in \{0,1\}^{n \times K}$ be the \textit{community membership matrix}, whose entries $Z_{ik}=1$ if and only if $g_i=k$. The association and cut of the partition $P$ are defined as \cite{von_luxburg_tutorial_2007}:
\vspace{-0.5em}\begin{align}
    \assoc{G}(\nvec{Z}) & = \sum_{i<j}^n A_{ij} \delta_{g_ig_j} = \frac{1}{2}\Tr(\nvec{Z}^T\nvec{A}\nvec{Z}), \label{eq:assoc} \\
    \cut{G}(\nvec{Z}) & = \sum_{i<j}^n A_{ij} (1-\delta_{g_ig_j}) = \frac{1}{2}\Tr(\nvec{Z}^T\nvec{L}\nvec{Z}).\label{eq:cut}
\end{align}

\vspace{-1em}
\subsection{Degree Corrected SBM (DCSBM)}
SBM was first proposed in social sciences as a random network model with community structure, where each node belongs to a community and edges between nodes are drawn independently based on their community membership \cite{holland1983stochastic,goldenberg2010survey}. The model is parameterized with community assignment vector $\nvec{g}$ and an edge probability matrix $\nmat{\theta}\in [0,1]^{K\times K}$ where $\theta_{kl}$ is the probability of an edge between the $k$th and $l$th communities and $K$ is the number of communities. The edge between nodes $i$ and $j$ is drawn from a Bernoulli distribution  with probability $\theta_{g_ig_j}$.  SBM has been used for inferring communities by maximizing the likelihood function of the observed network with respect to $\nvec{g}$   \cite{snijders_estimation_1997}. 

In \cite{karrer_stochastic_2011}, it is observed that network inference with SBM can result in erroneous community assignments when the degrees of the nodes are not uniformly distributed. In order to overcome this problem,  \textit{degree-corrected} SBM (DCSBM), in which degrees of nodes are  used in determining the probability of edge formation, has been proposed. This is done by assuming that the edge between nodes $i$ and $j$ comes from a Poisson distribution with mean $\lambda_{ij} = d_i d_j \theta_{g_ig_j}$. DCSBM leads to the following likelihood function, which can be maximized with respect to $\nvec{g}$ to find community structure:
\begin{align}
\label{eq:likelihood}
    \P(\A|\nvec{g};\ \nmat{\theta}) = \prod_{i<j}^n \frac{(\lambda_{ij})^{A_{ij}}e^{-\lambda_{ij}}}{A_{ij}!}. 
\end{align}

\vspace{-1.5em}
\section{Method}
\label{sec:method}

\vspace{-0.5em}
\subsection{Dynamic DCSBM}
\label{ssec:dynamicdsbm}

Recently, DCSBM has been extended to dynamic networks in \cite{ghasemian_detectability_2016,pamfil_relating_2018,bazzi_generative_2016}, where the network at each time is a DCSBM and community assignment of any node $i$ at time $t$ is modelled to be the same as the community assignment at time $t-1$ with a \textit{copying probability} of $q^t$. We base our dynamic DCSBM on this prior work but with a different assumption about network dynamics. Let $\nvec{g}=\{\nvec{g}^1, \dots, \nvec{g}^T\}$ and  $\nvec{\theta}=\{\nvec{\theta}^1, \dots, \nvec{\theta}^T\}$ be sequences of community assignment vectors and edge probability matrices at each time point for a dynamic network $\G$, respectively. Moreover, we assume that there are $K$ communities at each time. Different from previous work, we define a sequence of copying probabilities, $\nvec{q} = \{\nvec{q}^2, \dots, \nvec{q}^T\}$, where the $k$th entry of $\nvec{q}^t\in [0, 1]^K$, $q_k^t$, is the probability of a node at time $t-1$ staying in the $k$th community. Thus, our model allows each community to have its own copying probability $q_k^t$. This is a reasonable assumption since each community may have its own evolutionary dynamics, such that some communities may grow with time while others may stay stationary across time \cite{rossetti_community_2018}. Next, community assignments of nodes are modelled as follows. If $g_i^{t-1}=k$, then we assume $g_i^t=k$ with probability $q_{k}^t$, otherwise $g_i^t$ is equal to one of the $K$ communities with uniform probability. Based on this,  $\P(g_i^t=l|g_i^{t-1}=k) = \pi_{il}^t$ where $\pi_{il}^t=q_k^t+\frac{1-q_k^t}{K}$ if $k=l$ and $\pi_{il}^t=\frac{1-q_k^t}{K}$ otherwise. Finally, the community transition probabilities are assumed to be independent across nodes. Therefore, the prior distribution $\P(\nvec{g})$ is: 
\vspace{-0.5em}\begin{align}
\label{eq:prior}
\begin{split}
    \P(\nvec{g}; \nvec{q}) = \P(\nvec{g^1}) \prod_{t=2}^T \P(\nvec{g}^t|\nvec{g}^{t-1}) = \prod_{i=1}^n \P(g_i^1) \prod_{t=2}^T \prod_{i=1}^n \pi_{ig_{i}^t}^t,
\end{split}
\end{align}
where $P(g_i^1)$ is the prior probability of community assignment of node $i$ at $t=1$ and it is assumed to be uniformly distributed, i.e. $P(g_i^1=k) = 1/K\, , \forall k={1, \dots, K}$.  

\vspace{-0.5em}
\subsection{Dynamic community detection}
In this section, we show how maximizing the posterior distribution of dynamic DCSBM  can be transformed into a trace maximization problem, which can be solved using spectral clustering algorithms. Using Eqs. \ref{eq:likelihood} and \ref{eq:prior}, the posterior distribution of a dynamic network $\mathcal{G}$ following dynamic DCSBM is written as:
\vspace{-0.5em}\begin{align}
\label{eq:posterior}
    \P(\A,\nvec{g};\nmat{\theta}, \nvec{q}) \hspace{-0.2em} = \hspace{-0.2em} \prod_{t=1}^T \prod_{i<j}^n \frac{(\lambda_{ij}^t)^{A_{ij}^t} e^{\minus\lambda_{ij}^t}}{A_{ij}^t!}
    \prod_{i=1}^n \frac{1}{K} \prod_{t=2}^T \prod_{i=1}^n \pi_{ig_{i}^t}^t ,
\end{align}
where $\lambda_{ij}^t = d_i^td_j^t\theta_{g_i^tg_j^t}^t$ and $d_i^t$ is the degree of node $i$ at time $t$. Let $\L(\nvec{g})=\log \P(\A,\ \nvec{g}; \ \nmat{\theta}, \nvec{q})$, which can be written as follows by ignoring the terms that do not depend on $\nvec{g}$:
\vspace{-0.5em}\begin{flalign}
\label{eq:logposterior}
\L(\nvec{g})  \hspace{-0.2em} = & \hspace{-0.2em} \sum_{t=1}^T  \hspace{-0.1em} \sum_{i<j}^n  \hspace{-0.1em}  [A_{ij}^t\log(\theta_{g_i^tg_j^t}^t)\minus d_i^td_j^t\theta_{g_i^tg_j^t}^t ]+ 
\hspace{-0.2em} \sum_{t=2}^T \sum_{i=1}^n \hspace{-0.1em} \log(\pi_{ig_{i}^t}^t), \hspace{-1.5em}&
\end{flalign}
where the first and second terms are the log-likelihood and log-prior, respectively. First, consider the log-prior term in (\ref{eq:logposterior}). For fixed nodes $i$ and $j$, and fixed $t$, let $g_i^t=k$ and $g_j^t=l$ where $k, l \in \{1,\dots, K\}$. Then, the sum of log-priors of nodes $i$ and $j$ at time $t$ is $\log(\pi_{ik}^t) + \log(\pi_{il}^t) = \log(\pi_{ik}^t\pi_{il}^t)$. Due to independence, $\pi_{ik}^t\pi_{il}^t$ is the joint probability of nodes $i$ and $j$ being in communities $k$ and $l$ at time $t$, respectively. As community labels are arbitrary, it is more meaningful to quantify the joint probability of any two nodes being in the same community rather than the probability of individual nodes being in a particular community. Therefore, the joint probability $\pi_{ik}^t \pi_{jl}^t$ considered to be one of two values, namely when $i$ and $j$ are in the same community or in different communities:
\begin{align*}
    \pi_{ik}^t \pi_{jl}^t = \begin{cases} 
      p_{ij}^t/K, & k=l, \\
      (1-p_{ij}^t)/(K(K-1)), & k \neq l,   
   \end{cases}
\end{align*}
where $p_{ij}^t$ is the probability of nodes $i$ and $j$ being in the same community at time $t$ and the denominators are the normalization terms. Note that, $p_{ij}^t$ can also be calculated as $\sum_{k=1}^K \pi_{ik}^t\pi_{jk}^t$. Then, $\log(\pi_{ik}^t \pi_{jl}^t) = \log(p_{ij}^t/K)\delta_{kl} + \log((1-p_{ij}^t)/(K(K-1)))(1-\delta_{kl})$. This expression corresponds to the log-prior for a fixed node pair $(i,j)$ at time $t$. In order to write the log-prior term as a quadratic expression similar to spectral clustering, we add up the terms $\log(\pi_{ig_i^t}^t)$ for a fixed time $t$ as many times as necessary to generate terms for all node pairs (for $i<j$, we only need pairs in the form of $(i,j)$). This implies that we need $(n-1)$ number of log-prior terms for each node in  (\ref{eq:logposterior}) for a fixed time $t$. Thus, the second term of (\ref{eq:logposterior}) at time $t$ can be written by ignoring the terms that do not depend on $\nvec{g}$: 
\vspace{-0.5em}\begin{flalign}
\label{eq:priorsum}
    \frac{n\minus1}{n\minus1} \hspace{-0.2em} \sum_{i} \log(\pi_{ig_i^t}^t) \hspace{-0.2em} & = \hspace{-0.2em} \hspace{-0.2em} \sum_{i<j}^n \left[\frac{\log(p_{ij}^t)}{n\minus 1}\delta_{g_i^tg_j^t} \hspace{-0.2em} + \hspace{-0.2em}
    \frac{\log(1\minus p_{ij}^t)}{n\minus 1}(1\minus\delta_{g_i^tg_j^t})\right]. \hspace{-1.5em} & 
\end{flalign}
Next, we consider the log-likelihood term in (\ref{eq:logposterior}). At time $t$, we assume $\nvec{\theta}^t$ to be a planted partition model, i.e., $\theta_{kl}^t = \theta_{i}^t \delta_{kl} + \theta_{o}^t (1-\delta_{kl}) = (\theta_{i}^t - \theta_{o}^t)\delta_{kl} + \theta_{o} $, where $\theta_{i}$ is the intra-community connection probability and $\theta_{o}$ is the inter-community connection probability. Inserting this into the log-likelihood by ignoring the terms that do not depend on $\nvec{g}$:
\vspace{-0.5em}\begin{align}
\label{eq:likelihoodsum}
    \sum_{i<j} \{A_{ij}^t & (\log(\theta_{i}^t) - \log(\theta_{o}^t)) - d_i^t d_j^t (\theta_{i}^t - \theta_{o}^t) \} \delta_{g_i^tg_j^t}, \nonumber \\
    & =  \sum_{i<j} \beta^t A_{ij}^t \delta_{g_i^tg_j^t} - \gamma^t d_i^t d_j^t \delta_{g_i^tg_j^t},
\end{align}
where $\gamma^t = \theta_{i}^t - \theta_{o}^t$ and $\beta^t = \log(\theta_{i}^t) - \log(\theta_{o}^t)$. It is easy to see that (\ref{eq:priorsum}) and (\ref{eq:likelihoodsum}) are now similar to (\ref{eq:assoc}) and (\ref{eq:cut}), thus they can be written using a trace operator. Defining two matrices $\nmat{P}^t$ and $\nmat{Q}^t\, \forall t$ with entries $P_{ij}^t = P_{ji}^t = \log(p_{ij}^t)$ and $Q_{ij}^t = Q_{ji}^t = \log(1\minus p_{ij}^t)$, respectively, the log-posterior can be written as:
\vspace{-0.5em}\begin{align}
\label{eq:posteriortrace}
\begin{split}
    \L(\nvec{\nmat{Z}}) & \hspace{-0.1em} = \hspace{-0.1em} \sum_{t=1}^T \beta^t \Tr({\nmat{Z}^t}^T\nmat{A}\nmat{Z}^t) - \gamma^t \Tr({\nmat{Z}^t}^T{\nvec{D}^t}\nmat{Z}^t{\nmat{Z}^t}^T{\nvec{D}^t}\nmat{Z}^t) \\ & + \sum_{t=2}^T \frac{1}{n-1} \Tr({\nmat{Z}^t}^T(\nmat{P}+\nmat{L}_Q)\nmat{Z}^t),   
\end{split}
\end{align}
where $\nvec{L}_Q = \nmat{D}_Q - \nmat{Q}$ and $\nmat{D}_Q$ is a diagonal matrix with entries ${\nmat{D}_Q}_{ii} = \sum_{j=1}^n Q_{ij}$. 

\subsection{Constrained Dynamic Spectral Clustering}
\label{ssec:cdsc}
Maximizing (\ref{eq:posteriortrace}) with respect to $\nvec{Z}$ reveals the community structure of the dynamic network $\mathcal{G}$. As in spectral clustering, this problem is NP-hard since $\nvec{Z}$ is a binary matrix. Therefore, we relax $\nvec{Z}$ to take on any real value while imposing size constraints ${\nvec{Z}^t}^T\nvec{D}^t\nvec{Z}^t=\nvec{I},\ \forall t$.  Due to the constraint, the second term in (\ref{eq:posteriortrace}) becomes a constant, thus can be ignored during optimization. Thus, CD in a dynamic network $\G$ can be written as the following optimization problem: 
\begin{align}
    \label{eq:cdsc}
    \nmat{Z}^* = & \arg\max_{\nmat{Z}} \hspace{-0.1em} \sum_{t=1}^T \beta^t\Tr({\nmat{Z}^t}^T\nmat{A}^t\nmat{Z}^t) \hspace{-0.1em} + \hspace{-0.1em} \sum_{t=2}^T \frac{\Tr({\nmat{Z}^t}^T(\nmat{P}^t\hspace{-0.1em}+\hspace{-0.1em}\nmat{L}_Q^t)\nmat{Z}^t)}{n\minus1} \nonumber \\
    & \text{subject to } {\nvec{Z}^t}^T\nvec{D}^t\nvec{Z}^t=\nvec{I},\ \forall t.
\end{align}
This optimization problem is similar to EvoSC, where at each time point the first and second terms correspond to the snapshot and temporal costs, respectively. However, unlike EvoSC, our objective function is based on normalized association and the temporal cost  is a generalized version of temporal cost used in \textit{preserving cluster membership (PCM)}  \cite{chi_evolutionary_2007}. This is a generalization as we include copying probabilities into calculation of distance, whereas in PCM each community is assumed to evolve at the same rate.

The problem in (\ref{eq:cdsc}) can be solved via spectral clustering in an iterative fashion as follows. First, communities at $t=1$ can be obtained by static spectral clustering. Next, at any time $t>1$ a $K \times K$ matrix $\nmat{\Pi}^t = diag(\nvec{q}^t) + \frac{1}{K}(\nvec{1}-\nvec{q}^t)\nvec{1}^T$ is constructed where $diag(\cdot)$ is an operator that transforms a vector into a diagonal matrix and $\nvec{1}$ is a $K$-dimensional vector of ones. From $\nmat{\Pi}^t$ and $\nmat{Z}^{t-1}$, we calculate $\nmat{P}^t={\nmat{Z}^{t-1}}\log(\nmat{\Pi}^t{\nmat{\Pi}^t}^T){\nmat{Z}^{t-1}}^T$ and $\nmat{Q}^t={\nmat{Z}^{t-1}}\log(1\minus\nmat{\Pi}^t{\nmat{\Pi}^t}^T){\nmat{Z}^{t-1}}^T$ where logarithm is taken element-wise. Finally, spectral clustering is applied to the matrix $\beta^t\nmat{A}^t+\frac{\nmat{P}^t + \nmat{L}_Q^t}{n\minus1}$ with the constraint ${\nvec{Z}^t}^T\nvec{D}^t\nvec{Z}^t=\nvec{I}$. Since CD is performed individually at each time point , the number of communities can be  different at each time. Pseudo-code for the proposed approach is given in Algorithm 1.

\vspace{-0.6em}\subsection{Parameter Estimation}\vspace{-0.4em}
\label{ssec:parameterest}
The proposed method requires the estimation of copying probabilities $\nvec{q}$ and parameter $\beta^t$. These parameters are estimated in an iterative fashion similar to \cite{pamfil_relating_2018}. In particular, at each time $\nvec{q}^t$ and $\beta^t$ are randomly initialized with $\nvec{q}_0^t$ and $\beta_0^t$ and community structure is found as in Algorithm 1. Next, the community structure $\nvec{Z}^t$ is compared to $\nvec{Z}^{t-1}$ to update copying probabilities $\nvec{q}^t$.  $\nvec{Z}^t$ and $\nvec{A}^t$ are also used to compute $\theta_i^t$ and $\theta_o^t$ as in \cite{karrer_stochastic_2011} and $\beta^t=\log(\theta_i^t)-log(\theta_o^t)$. Lastly, $\nvec{Z}^t$ is updated by finding the community structure at time $t$ with the updated parameter values. This process is repeated iteratively $N$ times or till convergence. In our experiments, it was observed that copying probabilities and $\beta^t$ do not change after a couple of iterations.   

\begin{algorithm}[tb] \label{alg:cdsc}
\caption{Constrained Dynamic Spectral Clustering }
\textbf{Input:} Dynamic network $\mathcal{G}=(G^1, \dots, G^T)$, Number of communities $K^1, \dots, K^T$ \\
\textbf{Output:} Community Structure $P^*$
\begin{algorithmic}[1]
\FOR{t=1 to T}
\IF{t is equal to 1}
\STATE $\nmat{Z}^t \leftarrow$ Spectral clustering of $\nmat{A}^t$ with $K^t$ by \eqref{eq:assoc}.
\ELSE
\STATE Find parameters $\nvec{q}^t$ and $\beta^t$ as in Section \ref{ssec:parameterest}
\STATE $\nmat{\Pi}^t \leftarrow diag(\nvec{q}^t) + \frac{1}{K^t}(\nvec{1}-\nvec{q}^t)\nvec{1}^T$
\STATE $\nmat{P}^t\leftarrow {\nmat{Z}^{t-1}}\log(\nmat{\Pi}^t{\nmat{\Pi}^t}^T){\nmat{Z}^{t-1}}^T$ \STATE $\nmat{Q}^t\leftarrow{\nmat{Z}^{t-1}}\log(1\minus\nmat{\Pi}^t{\nmat{\Pi}^t}^T){\nmat{Z}^{t-1}}^T$
\STATE $\nmat{L_Q} \leftarrow \nmat{D}_Q-\nmat{Q}$
\STATE $\widehat{\nmat{A}}^t \leftarrow \beta^t\nmat{A}^t + \frac{1}{n+1}(\nmat{P}+\nmat{L}_{Q})$
\STATE $\nmat{Z}^t \leftarrow$ Spectral clustering of $\widehat{\nmat{A}}^t$ with $K^t$ by \eqref{eq:assoc}.
\ENDIF
\ENDFOR
\end{algorithmic}
\end{algorithm}

\vspace{-1em}
\section{Results}
\label{sec:results}
\vspace{-0.5em}
\subsection{Results for Simulated Networks} 
The performance of the proposed method is first evaluated on simulated networks and compared to state-of-the-art dynamic CD methods including PCM \cite{chi_evolutionary_2007}, DSBM \cite{xu_dynamic_2014} and GenLouvain \cite{pamfil_relating_2018}. First, we generate simulated networks based on Girvan-Newman (GN) benchmark networks \cite{girvan_community_2002}. At time point $t=1$, a GN network with 128 nodes divided into 4 equal sized communities is generated. For $1<t\leq T$, community assignments of each node is first determined by the copying probability $\nvec{q}^t = \nvec{q}\in[0,1]^4$, that is a node in the $k$th community at time $t-1$ stays in the $k$th community at time $t$ with probability $q_k$, otherwise it is randomly assigned to one of the 4 communities. For all time points, average degree and mixing coefficients are set to $16$ and $\mu$, respectively. 
Mixing coefficient $\mu$ indicates how noisy the community structure of the network is. The larger the $\mu$ is, the harder it is to detect the community structure. Comparison is done by calculating the normalized mutual information (NMI) \cite{danon_comparing_2005} for each method averaged over time and 50 Monte Carlo simulations.

\begin{figure}[b]
\vspace{-1em}
\begin{minipage}[b]{.49\linewidth}
 \centering
 \centerline{\includegraphics[width=\columnwidth]{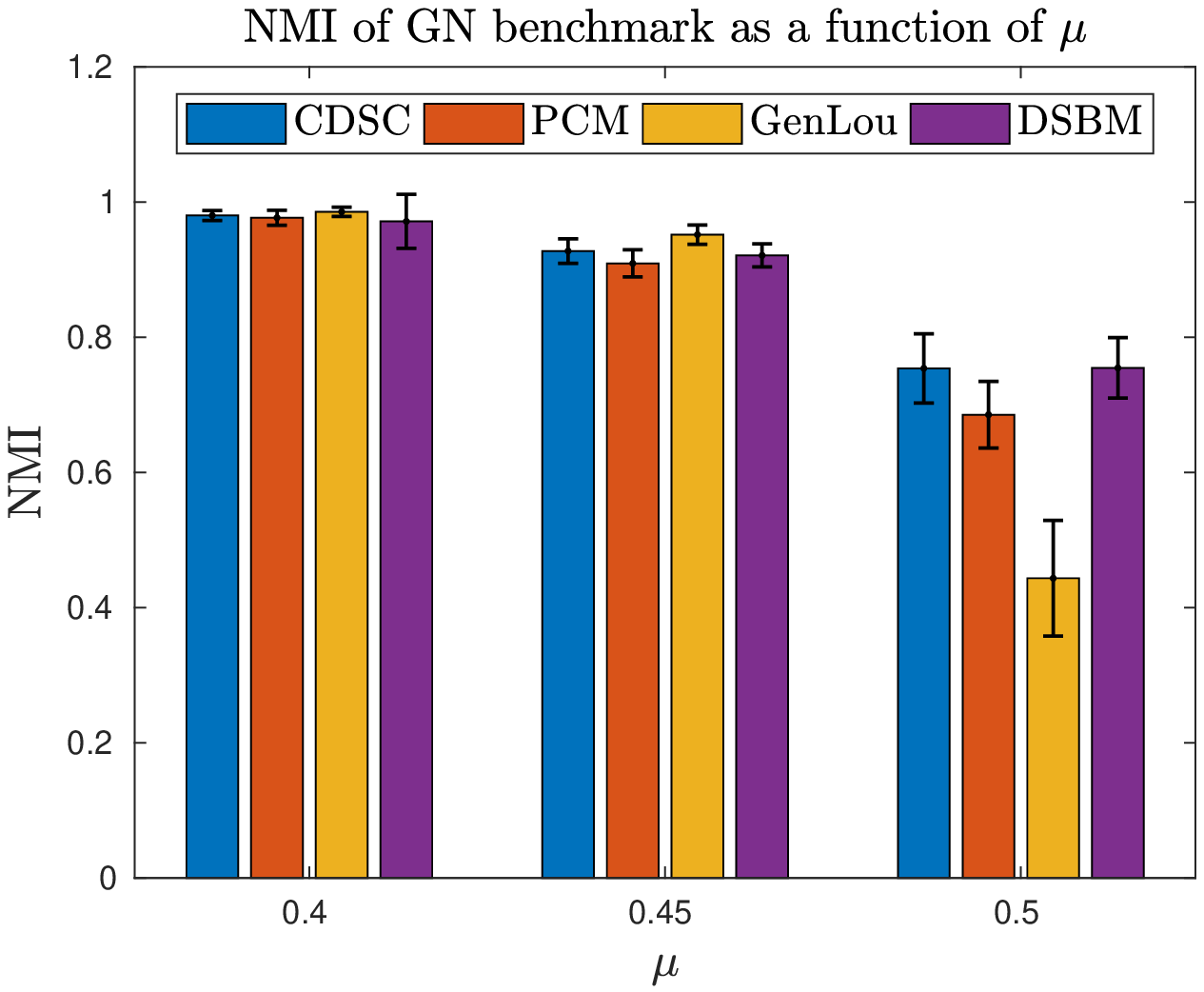}}
 \centerline{(a)}\medskip
\end{minipage}
\hfill
\begin{minipage}[b]{0.49\linewidth}
 \centering
 \centerline{\includegraphics[width=\columnwidth]{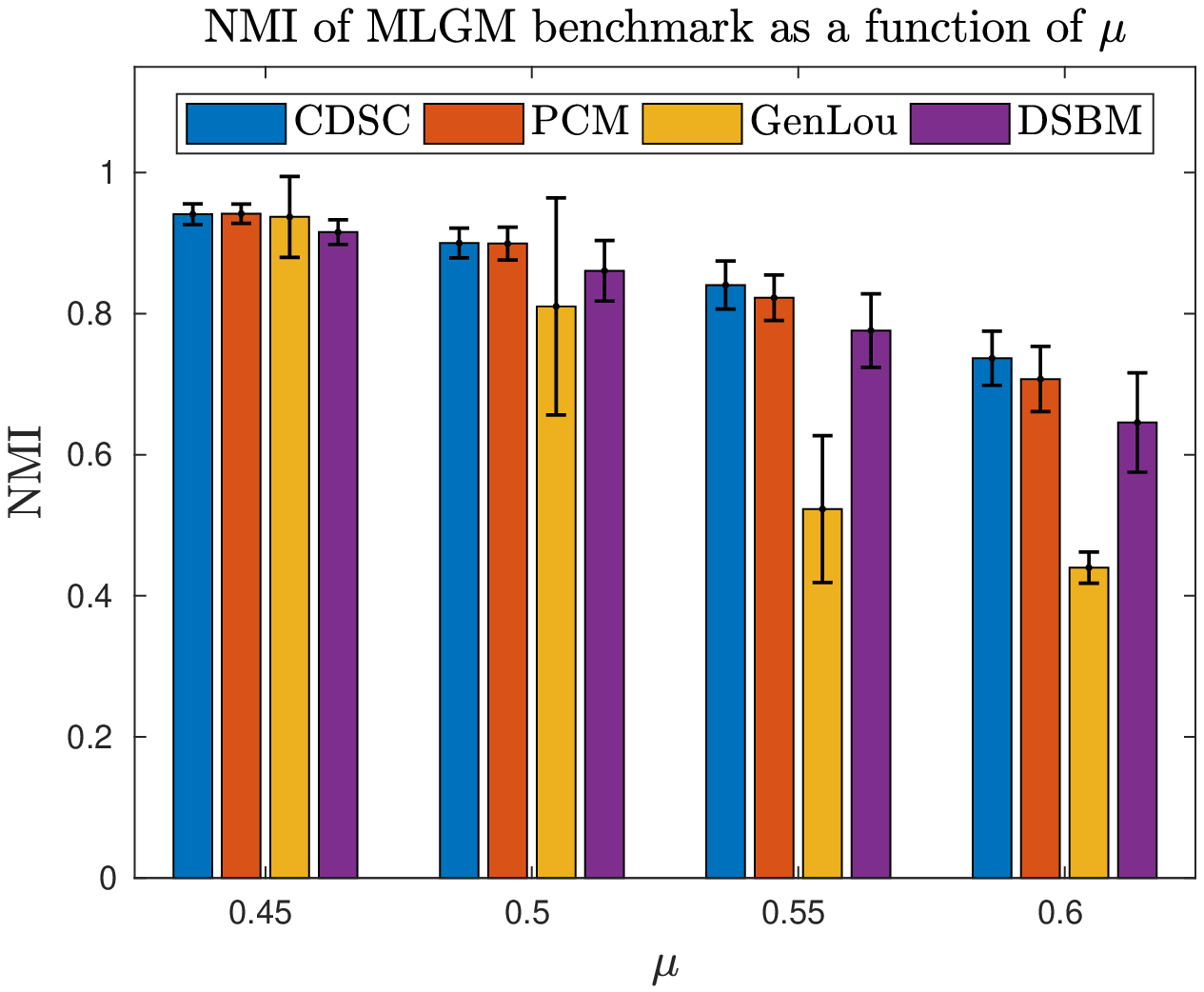}}
 \centerline{(b)}\medskip
\end{minipage}
\vspace{-1em}
\caption{Average NMI values for the different methods as a function of mixing parameter $\mu$: (a) GN benchmark networks and (b) MLGM benchmark networks.}
\label{fig:binarybenchmark}

\end{figure}

In Fig. \ref{fig:binarybenchmark}a, the results for GN benchmark can be seen for $\nvec{q} = [0.9, 0.6, 0.9, 0.6]$, T=10 and 3 different values of $\mu$. For PCM, the parameter $\alpha$ is set to 1 and $\beta$ is selected empirically between $0.1$ and $0.3$ as the one that gives the best normalized association value. Initial values of the parameters for GenLouvain and DSBM are set in a similar fashion as in the original papers \cite{pamfil_relating_2018, xu_dynamic_2014}. Finally, for all of the methods the number of communities are assumed to be known. For $\mu=0.40$, all algorithms yield high average NMI values as shown in Fig. \ref{fig:binarybenchmark}a, while the smallest variance in NMI is achieved by CDSC and PCM. As $\mu$ increases, the performance of all methods degrades. However, GenLouvain degrades faster than the others as seen in results for $\mu=0.50$, where the best result is achieved by CDSC both in terms of average NMI and variance across simulations. Finally, as CDSC is a generalized version of PCM, it always provides better accuracy than PCM (difference when $\mu=0.5$ is statistically significant at $\alpha=0.001$). The results indicate that incorporating copying probabilities that are dependent on community membership in DCSBM improves performance. 

The results given above indicate that CDSC provides higher accuracy than existing methods for GN networks. However, GN benchmark model is too simplistic in the way it generates the network as it does not account for heterogeneity in degrees and inter-community edge probabilities. For this reason, we evaluate the proposed method on a more complex benchmark proposed in \cite{bazzi_generative_2016}, referred as Multilayer Generative Model (MLGM) bechmark. This benchmark is generated using a dynamic DCSBM similar to the one mentioned in Section \ref{ssec:dynamicdsbm} and introduces heterogeneity in the degrees of nodes, community sizes, inter-community edge probabilities. Moreover, we modified the benchmark such that each community can have different copying probabilities.  The number of nodes is set to $128$, $T=10$ and copying probabilities are $\nvec{q}^t = [0.9, 0.6, 0.9, 0.6]$ for all $t$. At each time there are $4$ communities with different sizes and degrees of nodes are drawn from a power law distribution truncated between $8$ and $16$. Results are shown for four different values of $\mu$ in Fig. \ref{fig:binarybenchmark}b. For small values of $\mu$, all methods have similar NMI values. As $\mu$ increases, the proposed method performs the best giving the highest average NMI (difference between PCM and CDSC when $\mu=0.55$ and $\mu=0.6$ are statistically significant at $\alpha=0.001$).
\vspace{-1em}

\subsection{Results for Primary School Temporal Networks (PSTN):} The proposed method is applied to a real dynamic social network that depicts the connectivity between students and teachers in a primary school. The data is collected in October 2009 for one day using wearable sensors that measure face-to-face proximity. Temporal resolution of the data is 20 seconds, and there are 232 students and 10 teachers. The school is in session between 8:30 a.m. and 4:30 p.m. with two 20-25 minutes breaks at 10:30 a.m. and 3:30 p.m. and lunch time between 12:00 p.m. to 2:00 p.m. \cite{stehle2011high}. The raw data are divided into 13 minute intervals and a binary network is generated for each interval by connecting two individuals if they interact in the given time interval. The resulting dynamic network has $T=40$ time points and 242 nodes. 

\begin{figure}[b]
\vspace{-1em}
\begin{minipage}[b]{.49\linewidth}
 \centering
 \centerline{\includegraphics[width=\columnwidth]{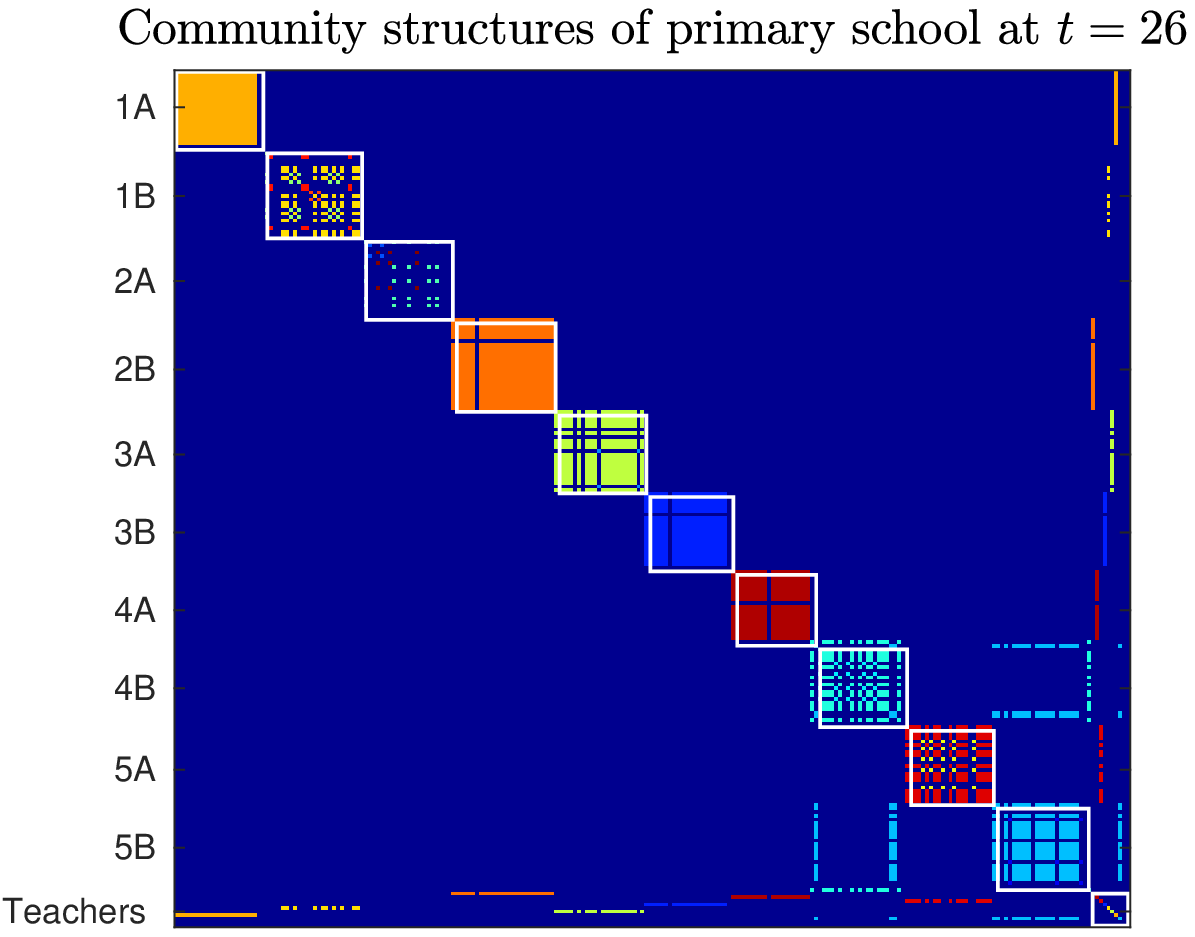}}
 \centerline{(a)}\medskip
\end{minipage}
\hfill
\begin{minipage}[b]{0.49\linewidth}
 \centering
 \centerline{\includegraphics[width=\columnwidth]{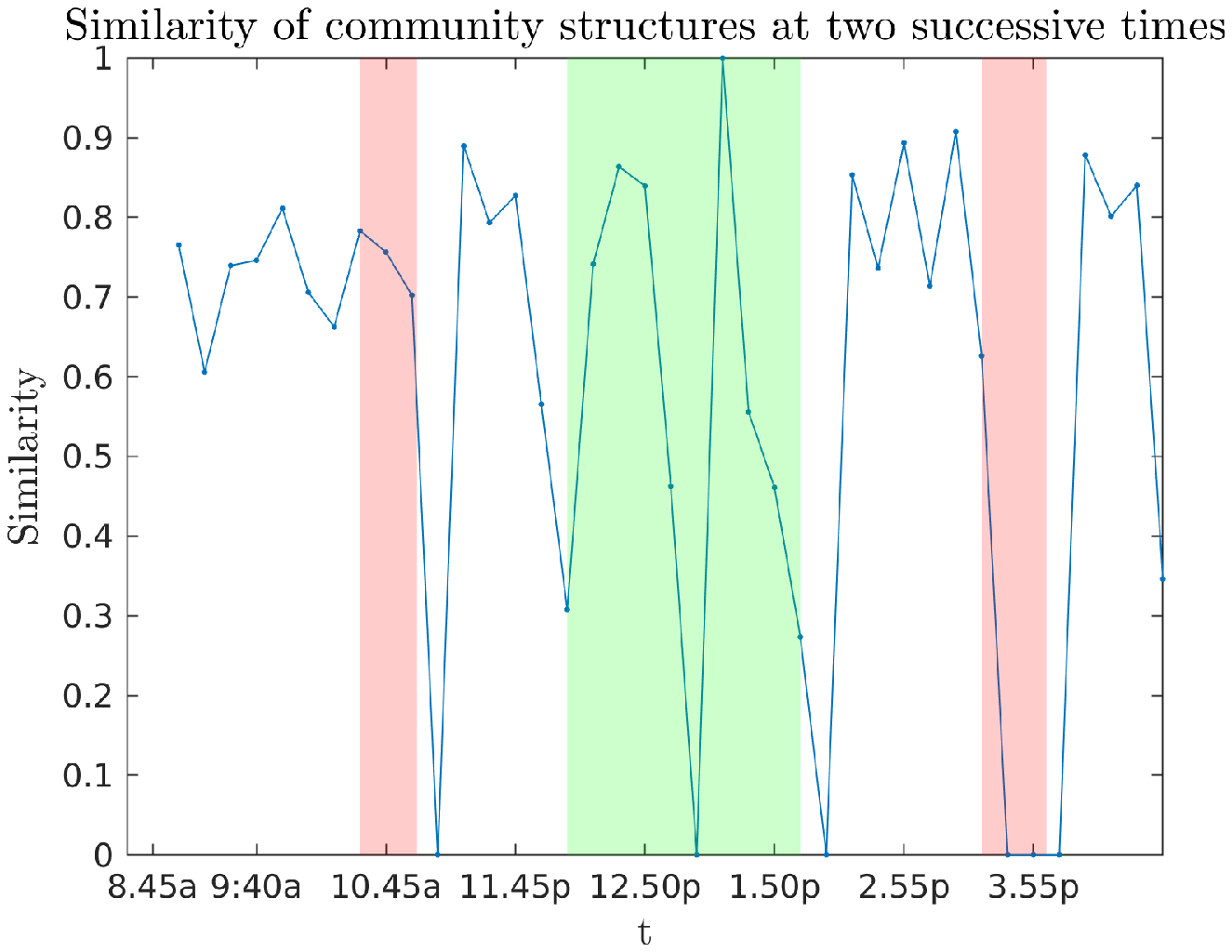}}
 \centerline{(b)}\medskip
\end{minipage}
\vspace{-1em}
\caption{(a) Detected communities for primary school data when the students are in classes. White rectangles correspond to different classes with the last rectangle corresponding to the teachers; (b) Similarity of community structures between consecutive time points, where red regions correspond to the two breaks and the green one to lunch time.}
\label{fig:primaryschool}
\end{figure}

The proposed method is applied to the constructed network where the number of communities at each time is selected as the number that maximizes \textit{asymptotic surprise} \cite{traag2015detecting}. In Fig. \ref{fig:primaryschool}a, the community structure of a time interval (between 2.15 p.m.and 2.30 p.m.) when students are in classes is shown as an example to indicate the effectiveness of the proposed method in detecting the communities. Fig \ref{fig:primaryschool}b shows the similarity between the community structures at consecutive time points, where the similarity is quantified by the weighted average of copying probabilities. It can be seen that the similarity is high for most times except during breaks and lunch time. These results agree with our intuition since students from different classes interact  with each other during breaks and lunch time resulting in a change in the community structure. Fig. \ref{fig:primaryschool}b also illustrates that the  proposed parameter estimation method described in Section \ref{ssec:parameterest} gives meaningful results.

\vspace{-0.8em}
\section{Conclusions}
\vspace{-0.4em}
In this work, a new algorithm for dynamic CD is introduced based on the equivalence between statistical network inference and spectral clustering. We first introduced a novel dynamic DCSBM that accounts for the differences in the evolutionary dynamics of different communities. We then proved the equivalency between statistical inference under this model and constrained spectral clustering for the planted partition model. Our derivation extends previous works that relate statistical inference and heuristic quality function optimization to dynamic networks. Moreover, the proposed method has been shown to be a generalization of PCM framework in EvoSC. Future work will exploit this relationship to analyze the consistency and scalability of the proposed algorithm and parameter estimation. 



\bibliographystyle{IEEEbib}
\bibliography{apos_library}

\end{document}